\documentclass[12pt]{article}
\usepackage{graphicx}
\begin{document}

\title{Dipole excitons in coupled quantum wells: toward an equilibrium exciton condensate}

\author{David W. Snoke\\Department of Physics and Astronomy, University of Pittsburgh\\ Pittsburgh, PA 15260, USA \\
snoke@pitt.edu}

 \maketitle

 \begin{abstract}
In recent years, experiments by several groups have demonstrated spontaneous coherence in polariton systems, which can be viewed as a type of nonequilibrium Bose-Einstein condensation. In these systems, the polariton lifetime is longer than, but not much longer than, the polariton-polariton scattering time which leads to the thermalization. By contrast, over the past twenty years several groups have pursued experiments in a different system, which has very long exciton lifetime, up to 30 microseconds or more, essentially infinite compared to the thermalization time of the excitons. Thermal equilibrium of this type of exciton in a trap has been demonstrated experimentally. In this system, the interactions between the excitons are not short-range contact interactions, but instead are dipole-dipole interactions, with the force at long range going as $1/r^{3}$. Up to now there has not been a universally accepted demonstration of BEC in this type of system, and the way forward will require better understanding of the many-body effects of the excitons.  I review what has been learned and accomplished in the past two decades in the search for an equilibrium BEC in this promising system.
\end{abstract}

\section{Introduction: Equilibrium and Quasi-equilibrium condensates} 

For many who are familiar with cold atom gases, the idea of a condensate of quasiparticles such as excitons or polaritons seems doubtful.  Often, a brief reading of solid state physics texts give the impression that a quasiparticle is not a ``real'' particle like an atom, because in many systems the number of quasiparticles is not conserved, being determined instead by the chemical potential. 

While that is true for some quasiparticles, there are two types of quasiparticle systems that do not fit this intuition. One type is a system in which stable quasiparticles are constructed out of other particles. Cooper pairs in superconductors and fractionally charged quasiparticles in two-dimensional electron gases are examples of stable quasiparticles that do not decay. Another type is a system in which metastable quasiparticles are constructed out of excitations of an underlying system; their total number is determined not by spontaneous thermal excitations but by an external pump that generates a population of particles on demand. 

In the first type of system, it is easy to see that stable, permanent bosonic quasiparticles can undergo condensation. Cooper pairs of electrons are well known to become a condensate in the superconducting phase transition \cite{snokebook,leggett}.  In the second type of system, Bose condensation is also possible. The criterion for condensation in this case is that the lifetime of the metastable quasiparticles must be long compared to their thermalization time, which can be determined either by the particle-particle collision time or by the time scale of their coupling to an external bath, e.g. phonon emission and absorption.  We can experiments in two different limits for the excitation conditions when the lifetime is long compared to the thermalization time. In a time-dependent experiment, particles are generated in a very short time by an external pump, then reach thermal equilibrium among themselves, and finally the number of particles decays slowly, maintaining thermal equilibrium the whole time. In a steady-state experiment, a continuous pump generates a number of particles at each moment which is small compared to the total population, enough to replace the small number which decays at each point in time. 

In each of these two types of experiments, if the ratio of the lifetime to the thermalization time approaches infinity, then obviously the physics will be no different from that of a population of permanent atoms in equilibrium.  If the lifetime is just a bit longer than the thermalization time, however, then condensation can occur but important nonequilibrium effects can come in to play, which are the subject of this book. The absolute time scale does not matter, on the ratio of time scales.

The polariton condensates \cite{science,deveaud,vina,vort}, which are the subject of other articles in this book, typically have lifetime just a few times longer than the thermalization time of the particles. There is another quasiparticle system which has been extensively studied since the early 1990s, however, which has lifetime much longer than the thermalization time.  Semiconductor heterostructures can be designed which allow control of the exciton lifetime and the exciton-exciton interactions.

\section{Excitons in coupled quantum wells}
\label{cqw1}

\begin{figure}
\begin{center}
\includegraphics[width=0.35\textwidth]{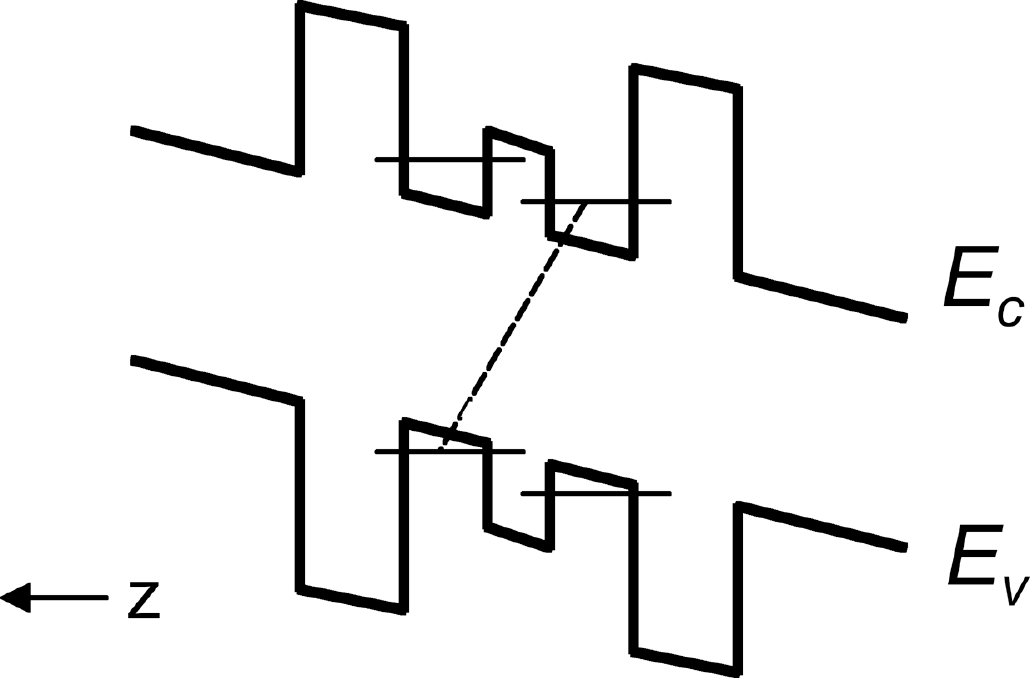}
\end{center}
\caption{Schematic band diagram of the coupled quantum well structure in an electric field. The heavy solid lines indicate the bands, while the thin solid lines indicate the quantized states in the wells. The indirect excitons are created from electrons in the lowest conduction band state and holes from the highest valence band state, as indicated by the dashed line.}
\label{schem}
\end{figure}

Nearly as soon as the technology for quantum heterostructures had been developed in the 1970s, an early paper by Lozovik and Yudson\cite{yudson} predicted a new type of excitonic Bose condensate based on coupled quantum wells. Figure \ref{schem} shows the band diagram of a typical structure in the presence of electric field. A variation uses a type-II structure in which the excitons are indirect in both real space and momentum space. In each case, the excitons can move freely in two dimensions, while they are confined in the perpendicular direction. There are several appealing properties of this system in regard to the search for excitonic condensation.

First, the excitons in this system can have very long lifetime, up to tens of microseconds \cite{snoke-diffprl}. As illustrated in Fig.~\ref{schem}, when an electric field is applied normal to the wells, excited electrons and holes are pulled into adjacent layers. To recombine, the electron and hole must find each other through a tunneling barrier. Since the spatial overlap of the exciton and hole wavefunctions in the barrier is exponentially suppressed inside the tunneling barrier, the recombination rate can be up to six orders of magnitude lower than the intrinsic exciton lifetime, for typical structure sizes and typical electric fields \cite{szy}. Enhancement of the exciton lifetime in this type of structure was first shown by Mendez and coworkers in the later 1980's \cite{mendez}. Since the relevant thermalization times for excitons are picoseconds to nanoseconds, both for exciton-phonon scattering \cite{ivanov} and for exciton-exciton scattering at typical densities (see below), there is no issue of reaching local equilibrium. There can, however, be issues of global equilibrium and localization in a disordered potential, as discussed below.

\begin{figure}
\begin{center}
\includegraphics[width=0.45\textwidth]{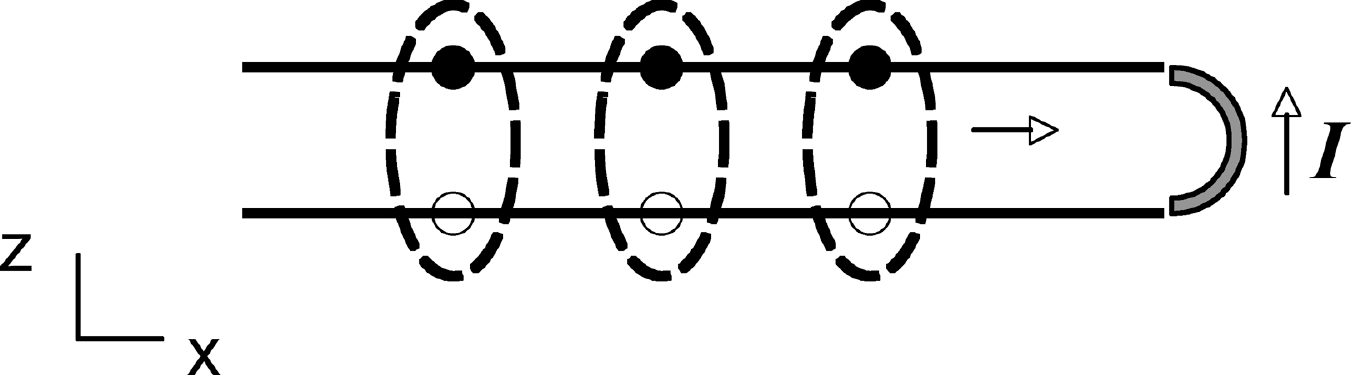}
\end{center}
\caption{Schematic of current flow when excitons move in a coupled quantum well system. As excitons move to the right, a current flows upward from bottom to top well, if there is a conducting connection.}
\label{flow}
\end{figure}

A second appealing characteristic of this system is that the interaction of the excitons is entirely repulsive when the spatial separation between the electron and hole layers is high enough; as indicated in Fig.~\ref{flow}, the excitons interact as aligned dipoles.  Lozovik and Berman \cite{lozo-berm}, using a mean-field theory, found that the overall interaction is repulsive for spatial separation greater than about three times the exciton Bohr radius. This is in contrast to excitons in single quantum wells, which typically have an attractive interaction for some separations and some relative spin states, which gives rise to complicated spectra which include biexcitons (excitonic molecules), and in some materials also electron-hole liquid \cite{EHL}. In typical coupled quantum well experiments, the indirect exciton line is a single line, and there is no biexciton line or any other line nearby in energy which might complicate the analysis.  An overall repulsive interaction is also important for stabilizing a condensate of the excitons.

Another appealing feature, pointed out by Lozovik and Yudson \cite{yudson}, is that a current of excitons in a two-dimensional plane corresponds to a real current, if a contact is made between the top and bottom layers, as shown in Fig.~\ref{flow}. Thus, a superfluid flow of an excitonic condensate would also be a Schafroth superconductor \cite{schaf}. 

\begin{figure}
\begin{center}
\includegraphics[width=0.6\textwidth]{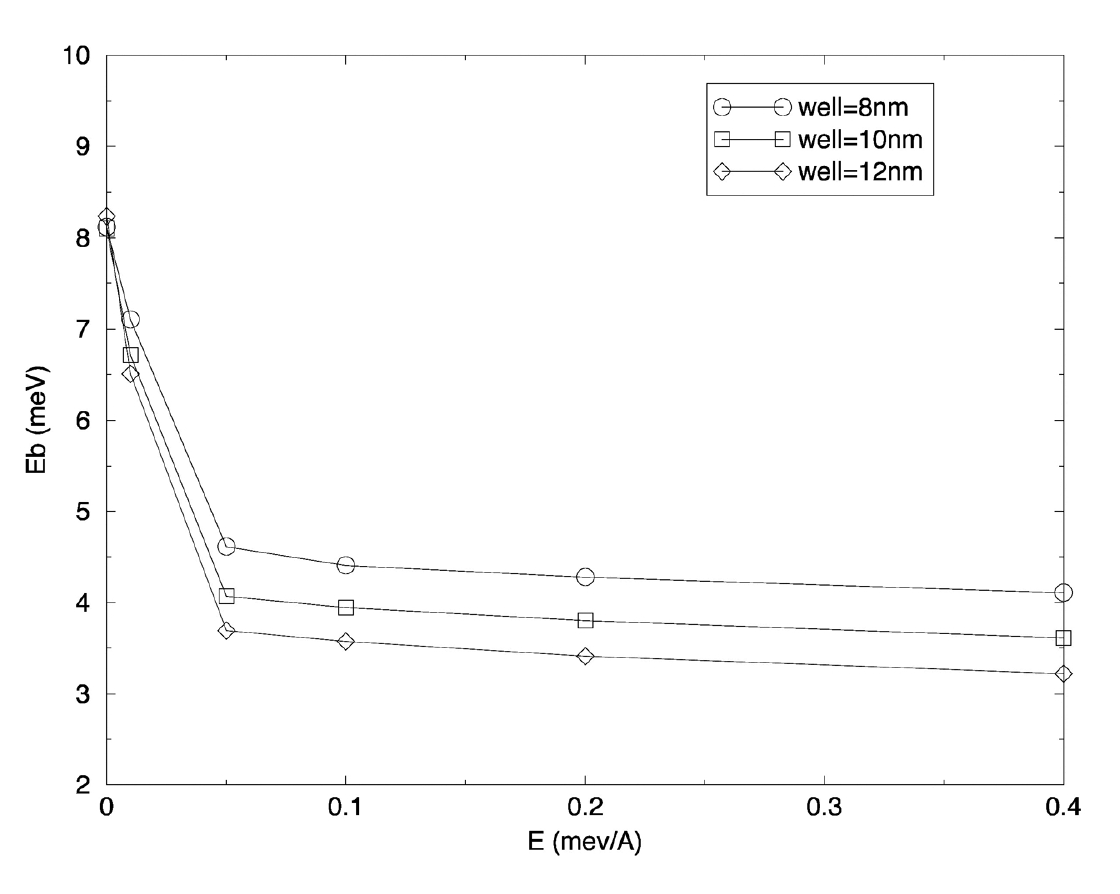}
\end{center}
\caption{Binding energy of excitons in coupled quantum well structures as a function of applied electric field. From Ref.~\protect\cite{szy}.}
\label{szy}
\end{figure}

Excitons in coupled quantum wells also can exist at high temperatures, especially high when compared to atomic gases at nano-Kelvin temperatures. The binding energy of excitons in single narrow GaAs quantum wells is around 10 meV. When electric field is applied to separate the electron and hole spatially, this binding energy goes down, but because there are outer barriers (see Fig.~\ref{schem}), the electron and hole do not keep getting further away from each other, and therefore the binding energy reaches a limiting value, as shown in Fig.~\ref{szy}. In GaAs coupled quantum wells the binding energy is typically around 4 meV \cite{szy}.This means that excitons in GaAs structures can persist, coexisting with free electrons and holes, all the way up to 90 Kelvin or so \cite{nature,ssc-exion}, depending on the total carrier density.  In other semiconductors, excitons can be much more deeply bound, e.g. excitons in II-VI materials typically have binding energies of 10-30 meV \cite{IIVI}; GaN excitons have binding energy of 27 meV and ZnO excitons have binding energy of 60 meV \cite{gan}. It is therefore not a stretch to expect an excitonic condensate at room temperature. However, deeper binding energy means smaller excitons (see Ref.~\cite{snokebook}, Section 2.3), which means that the excitons are much more sensitive to local disorder; so far, experiments on free excitons (not polaritons) in quantum wells in materials other than GaAs show large effects of disorder and have not made much progress toward equilibrium BEC. Exciton experiments in GaAs are typically done at temperatures of 1-10 K, so that $k_BT$ is well below the excitonic binding energy. 

\section{Past claims of exciton BEC in coupled quantum wells}
\label{cqw2}

After relatively long lifetime of excitons had been demonstrated in coupled quantum wells in the late 1980's, experiments proceeded immediately to try to observe excitonic condensation. The earliest claim was by Fukuzawa and coworkers \cite{fuku} in 1990.  The claim was based on the observation of narrowing of the exciton luminescence spectrum; the spectral width dropped sharply at low temperature and high electric field.  However, a year later, the same group \cite{kash} reported that the spectra could be better interpreted by a Fermi-Dirac distribution for the excitons. Although all excitons are bosons, it is well known that in a disordered potential, the exciton spatial distribution can be mapped to a Fermi-Dirac energy distribution. In essence, in a disordered landscape, one energy corresponds to one spatial location. If the particles repel each other, then only one particle can fit in one spatial location. Thus, only one particle can occupy one energy level, just as in Fermi-Dirac statistics. Evidence that this was occurring was seen in the fact that the spectrum shifted to lower energy at the density decreased, and the fact that the width of the spectrum was comparable to $k_BT$.  In this interpretation, the narrowing of the spectrum at low temperature occurred because the density of exciton states varied rapidly in the spectral range of interest, and therefore, as the temperature dropped, the width of the energy distribution dropped nonlinearly. 

This experiment points out an important historical milestone. Although excitons have been studied in quantum wells since the 1970s, it was not until the past decade that the characteristic energy fluctuations due to disorder could be made small compared to the exciton binding energy. In order to prevent exciton ionization, the temperature must be kept well below 100 K in GaAs structures, i.e., much less than the exciton binding energy. At these low temperatures, most early GaAs quantum well structures had exciton energy fluctuations due to disorder which were comparable to or larger than $k_BT$.  The luminescence line width of the excitons at low temperature is a rough measure of the magnitude of the fluctuations of the local potential due to disorder. (The broadening of the spectral line width due to disorder is known as ``inhomogeneous broadening,'' in contrast to ``homogeneous broadening'' due to intrinsic scattering processes, discussed below.) In most experiments before the year 2000 and in many since, excitons in quantum wells were mostly sitting trapped in local disorder minima, not moving at all. Clear evidence of long-range exciton motion in quantum well structures was first reported in 2004 \cite{kotthaus,snoke-diffprl}, in structures with inhomogeneous broadening around 0.3 meV (which corresponds to temperature about 4 K). High mobility of unpaired, free electrons and holes in quantum heterostructures was seen much earlier, e.g. in the famous quantum Hall experiments, because free carriers form a flat Fermi level as low-energy states in a disordered landscape are filled up, leading to a mobility edge, above which the electron and hole states are nearly translationally invariant. This does not generally occur with excitons, however, because when the carrier density is large enough to create a Fermi level, the exciton binding is screened out. 

Following these early experiments, Butov and coworkers, in a series of papers, reported evidence for exciton BEC in coupled quantum wells based on an increase of low-frequency intensity fluctuations \cite{butov-fluct}, an increasing diffusion constant when magnetic field was applied \cite{butov-diff}, an upturn in the luminescence intensity after a pump laser was turned off \cite{butov-cool}, enhanced luminescence in a localized spot \cite{butov-nature-trap}, and optical coherence seen in spots of exciton luminescence formed at the interface of an electron gas and a hole gas \cite{nature-butov,butov-coh}. 

Large low-frequency fluctuations of the indirect exciton luminescence were also seen by Timofeev and coworkers \cite{timo-fluct} and Krivolapchuk and coworkers \cite{kriv}. Low-frequency fluctuations of coupled quantum well excitons at high excitation density were found in at least some experiments to occur near the exciton-electron-hole plasma Mott transition \cite{fluct}, and could be interpreted as the result of fluctuating local electric fields which occur when the fraction of free electrons and holes fluctuates. A spectral narrowing of the photoluminescence from the excitons was also reported in these experiments, as in the IBM experiments, but no evidence of coherence was reported to occur at the point of strong fluctuations, and the emission spectrum was largely featureless, with no $\delta$-function-like peak. Temporal fluctuations were not reported in conjunction with later claims to exciton condensation by Butov and coworkers \cite{butov-coh}.

The increase of diffusion constant reported as evidence of condensation \cite{butov-diff} was deduced indirectly on the basis of the reduction of luminescence intensity lifetime for excitons created next to a mask, which was interpreted as arising from excitons moving under the mask. One problem with interpreting these data is that at high density and low lattice temperature, more than one process can lead to fast expansion of excitons away from the excitation spot. One such process is a ``phonon wind,'' in which hot nonequilibrium phonons created by the laser excitation can push the excitons \cite{wind}. Another effect is that the excitons, which strongly repel each other, have enough pressure to push each other outward more quickly at high density \cite{zoltan-equil}. The ambiguity of outward diffusion as evidence of superfluidity has led many experimentalists to turn to trapping the excitons (using the methods discussed below), instead of simply creating excitons in a spot where the laser hits the sample. In a trap, condensation should lead to {\em inward} motion of the excitons, as the excitons enter the ground state in the center of the trap, instead of an {\em outward} motion, which could be caused by phonon wind or exciton pressure.

Butov and coworkers also reported evidence of Bose statistics in the rapid cooling of excitons after a pump laser was turned off \cite{butov-cool}. This argument was based on fits to cooling rates computed for exciton-phonon emission \cite{ivanov}. Although the fit to the theory was good, the argument is somewhat indirect; an increase of the cooling rate is already expected just from the change of carrier temperature when the pump laser is turned off.  Another experiment \cite{butov-nature-trap} showed a bright spot which appeared at a fixed point on the sample and which narrowed spatially at low temperate; this was interpreted as a condensation of excitons in a local minimum in a disordered potential. One problem with the interpretation of this work is that at the time, the role of free charge carriers in the n-i-n doped structures \cite{snoke-ssc,rap-prl,lev-prl} was not well understood. In particular, it is now well understood that free current can tunnel through the heterostructures in filaments during these experiments \cite{butov-fil}.  This comes about because in all these coupled quantum well structures, when voltage is applied normal to the wells in order to make the excitons spatially indirect, as discussed above, a small current will flow through the structure, due to tunneling through the barriers. This current does not tunnel evenly \cite{laikht-tun}, because the tunneling rate depends exponentially on the barrier height. Therefore the current will mostly find a few individual points where the barrier height is slightly lower than the average. Bright spots of luminescence seen at fixed points on the samples are almost always due to this type of filament tunneling. 

Most recently, Butov and coworkers have reported interference patterns in the luminescence of excitons created at the interface of a hole gas and an electron gas \cite{butov-coh}.  This interface is seen in a dramatic ring pattern in the exciton luminescence \cite{nature,nature-butov,snoke-ssc,rap-prl,lev-prl}; it comes about because of competing effects of tunneling of free carriers through the barriers and hot carriers generated by the laser which hop over the barriers. A periodic modulation of the ring into small dots seen by Butov and coworkers has been explained as due to instabilities in the long-range Coulomb forces between the carriers \cite{cardiff}. The spectral width of the exciton luminescence reported in Ref.~\cite{butov-coh} was about 2 meV, which corresponds to a coherence time of less than a picosecond, much less than the lifetime of the excitons. It was argued by Butov and coworkers that the temperature dependence, in which both the interference fringes and the localized spots in the luminescence appeared only at low temperature, were consistent with an effect like Bose condensation which only occurs at low temperature. It should be noted that the reduction of the light source size will itself enhance the observation of coherence.

Timofeev and coworkers have also reported evidence for Bose-Einstein condensation in coupled quantum wells, in a series of papers \cite{tim1}. In these experiments, a mask was used to cover all but a tiny region of a coupled quantum well structure; this was done to restrict the observation to one local minimum in the (strongly) disordered potential. At low temperature, sharp peaks were seen in the exciton luminescence spectra. One interpretive difficulty with these experiments is that with such small traps, which are effectively quantum dots formed by minima in the disorder potential, the number of excitons that can accumulate in a single trap is very small. As mentioned above, the indirect excitons have a strong dipole-dipole repulsion, which means that a shallow trap will be quickly filled; for a trap with area of $(0.1~\mu\mbox{m})^2$, the best estimate of the energy shift \cite{zimm-corr}, $\Delta E = 0.1de^2n/\epsilon$, gives a blue shift of  about 15 meV for 10 excitons in the trap, for well separation $d = 100$ nm. The number of particles in a small trap can be kept low so that the energy shift is low, but when the number of particles is very small, it is hard to define what we mean by a condensate.

Despite a number of claims and intriguing experimental effects, there is no general consensus that Bose-Einstein condensation has been confirmed in observations of indirect exciton luminescence in coupled quantum wells. As discussed below, one of the reasons is the strong dipole-dipole repulsion of the excitons, which makes theoretical predictions of the conditions for condensation harder. Nevertheless, progress has been made in understanding the interactions and in trapping the excitons and seeing them reach equilibrium, and there is no in-principle reason why we should not expect to see Bose-Einstein condensation of excitons in equilibrium in this type of structure. 

\section{Equilibration, transport, and renormalization in traps}
\label{cqw3}

As discussed above, there is great advantage to confining the excitons in a trap, analogous to the optical traps for cold atoms, instead of just creating excitons with a laser and allowing them to expand freely out of the excitation region. In a trap, the condensate will have a telltale inward contraction to a spatially compact ground state; this has been seen with cold atom condensates \cite{ketterle} and polariton condensates \cite{snoke}. 

Two type of traps have been explored in depth. One method uses inhomogeneous tensile stress \cite{negoita} to shift the exciton energy bands.  The second uses electrostatic potential to shift the indirect exciton energy. Although the excitons are charge neutral, their energy depends on the applied field (see Fig.~\ref{szy}), in what is known as the quantum-confined Stark effect \cite{qcse1,qcse2}. Several groups have successfully shown that excitons can be trapped or manipulated by applied electric fields \cite{kotthaus-trap,rapaport-trap,butov-gate}. The stress method typically creates traps which are 10-50 microns wide, while the electrostatic method can create traps as small as a few microns. In general, both are limited to trap depths of a few tens of meV. In the case of the stress traps, the upper limit is due to shear stress which can break the sample, while in the electrostatic traps, an upper limit arises because high electric fields lead to in-plane forces that ionize the excitons. 

Recent work showed that indirect excitons reach equilibrium both energetically and spatially in stress-induced traps \cite{zoltan-equil}. Spatial equilibration in a trap requires that the diffusion length of the excitons be comparable to or greater than the size of the trap. The diffusion length is defined as $l = \sqrt{D\tau}$, where $D$ is the diffusion constant and $\tau$ is the particle lifetime. If it is much shorter than the trap size, the particles can equilibrate locally in momentum space but will not have a common chemical potential across the whole trap. Recent experiments \cite{snoke-diffprl,kotthaus} have demonstrated very long diffusion lengths, up to several hundred microns. At low temperature and low density, however, the excitons can become stuck in local minima in the disorder \cite{zoltan-stuck,zoltan-review}, and the diffusion length becomes much shorter than the trap size. 

There are therefore two limits on the density of excitons which can be generated in these structures. On one hand, there is a low-density limit. At low density, the critical temperature for condensation is low (nominally, the critical temperature is proportional to the temperature in a weakly interacting, trapped two-dimensional boson gas \cite{2dbec}). But if the temperature is low compared to the energy fluctuations due to disorder, then the excitons will become trapped in low-energy minima in the disorder potential, and will not act as a free gas. On the other hand, if the density is too high, the excitons can undergo a Mott transition to free electrons and holes \cite{ssc-exion}. Stern et al. \cite{stern} have 
shown evidence that indirect excitons can undergo a Mott transition at densities of the order of $2\times 10^{10}$ cm$^{-2}$; although they did not see a slow decrease of the exciton Rydberg, this is not inconsistent with the basic results of the theory of exciton ionization \cite{ssc-exion,manzke} which shows that a sudden transition to ionized electrons and holes can occur due to screening of the electron-hole Coulomb interaction at densities well below the absolute upper limit of $n \sim 1/a^2$, where $a$ is the exciton Bohr radius. 

Well below the density at which excitons become ionized, there can be another limit on the exciton density in traps. Because of the repulsive exciton-exciton interaction, there is a increase of the exciton chemical potential with increasing density, essentially a local potential energy increase due to the pressure from neighboring particles. This energy shift acts to cancel out the trapping potential and flatten it \cite{zoltan-review}. Thus, depending on the trap depth, the trap may be washed out by the interaction energy of the excitons.  

The interactions also lead to other complications. In general, the condensate fraction is depleted by interactions, which leads to a change of the critical temperature; the critical temperature for an interacting gas can be much less than that of an ideal gas at the same density. Laikhtman and Rapaport \cite{laikhtrap} have proposed that at high density, the dipole excitons in these structures no longer act as a gas but instead as a correlated liquid.  Their argument invokes reasonable approximations to conclude that the wave function of each exciton is highly localized by its nearest neighbors rather than extended like a plane wave. This does not mean that Bose condensation is impossible at high density--- liquid helium, after all, undergoes condensation as a liquid--- but it may mean that the critical temperature drops with increasing density rather than increasing. It may also mean that the canonical telltale for condensation in a weakly interacting Bose gas, namely a peak of the occupation number at $k=0$, may not be easily seen in these systems--- in liquid helium, the condensate fraction at $k=0$ is less than 10\%, and this peak cannot be easily seen in the momentum distribution \cite{sokol}. It may therefore be better to look for hydrodynamic effects of condensation of excitons at high density, such as quantized vortices or superfluidity, similar to the behavior of liquid helium.

All of this points out the importance of a proper calculation of the condensation phase transition of the excitons accounting for their interactions. Up to the present, there has been no fully self-consistent many-body calculation of the critical temperature and spatial distribution of the interacting dipole exciton gas in two dimensions. This calculation is much more complicated than the case of cold atoms for several reasons, none of which is related to the disorder or band structure of the solid. One difficulty is that the $1/r^3$ interaction of the excitons makes the effects of correlation much more important than in the weakly interacting gas \cite{zimm-corr}. Another difficulty is that there is exchange between both electrons and holes, which have comparable mass, makes the exciton-exciton interaction harder to calculate accurately \cite{laikht,zimm-corr,drummond-int,ceperley}. Progress has been made recently both experimentally and theoretically in determining the effects of the exciton-exciton interaction. A recent experiment \cite{zoltan-int} gives a measure of the interactions that does not depend on the calibration of the exciton density, a number which is notoriously hard to determine accurately. Theory by Zimmermann \cite{zimm-fit} fits the experimental data within a factor of 2 (which is an accomplishment, since mean-field theory is wrong by an order of magnitude), but gives a much flatter temperature dependence than seen in the data.

\section{Bilayer excitons, Coulomb drag and the pairing transition}

A variation of the coupled quantum well system which has received much attention recently is one in which a coupled quantum well system is given a permanent population of electrons via doping and gating. Since the population of carriers is permanent, transport measurements must be used to deduce the properties of the carrier population, rather than luminescence.
\begin{figure}
\begin{center}
\includegraphics[width=0.6\textwidth]{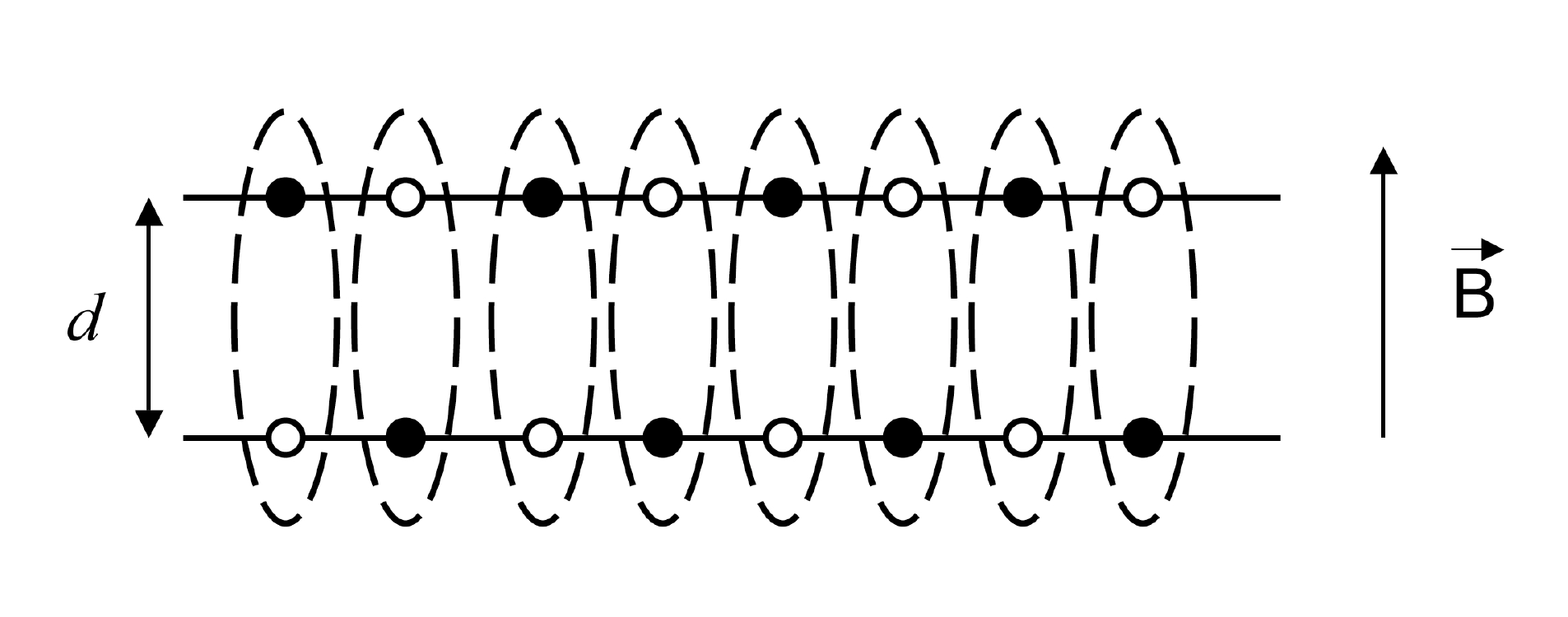}
\end{center}
\caption{Schematic of excitons in a bilayer system with a permanent population of electrons due to doping. The electrons each half-fill the lowest Landau level in a magnetic field.}
\label{bilayer}
\end{figure}

In a series of experiments performed by Eisenstein and coworkers \cite{eisen1} and Shayegan and coworkers \cite{tutuc}, a magnetic field was applied normal to the wells to put the electron gas into a state of filling factor of exactly $\nu=1$ for the lowest Landau level, which corresponds to filling factor $\nu=1/2$ for each of the two coupled wells separately. In this case, each layer could be viewed as having the lowest Landau level half filled with free electrons and half filled with free holes. ``Coulomb drag'' transport experiments showed that a current of carriers in one well was correlated with a voltage in the other well, which implied that the electons in one well were spatially correlated with holes in the other well, i.e. excitons, as illustrated in Fig.~\ref{bilayer}. 

The drag experiments show that the electrons and holes are spatially correlated with each other similar to Cooper pairs, but pairing by itself is not synonymous with condensation. In the case of BCS superconductors, pairing occurs at the same temperature as condensation, because the binding energy of the pairs is small compared to the $k_BT_c \sim n^{2/3}\hbar^2/m$ for condensation of stable pairs, where $n$ is the density and $m$ is the pair mass. In other words, when the pairs in a BCS superconductor form, they are instantly already at a temperature low compared to the critical temperature for condensation of stable pairs at that density.  This need not always be the case, however. If the pair binding energy is large, then the pairs can form stably at high temperatures, well above the temperature for condensation. In this case there will be two transitions: one for pairing at high temperature, and a second transition to condensation at a lower critical temperature.  This is the case for excitons are generated by optical excitation; the Coulomb attraction between the electrons and holes is so strong that the pairs are stable at temperatures much higher than the condensation temperature, and it also may be what happens in the high-$T_c$ superconductor ``pseudogap'' transition. 

In the case of the bilayer experiments of Eisenstein, Shayegan and coworkers, the ground state of the system at $T=0$ is essentially a BCS state, but with the Landau orbitals taking the role of the $k$-states in a standard BCS system. As $T$ increases, the condensate is depleted, but is still expected to exist at finite temperature with reduced coherence. Condensation would be directly shown by evidence of coherence; for example, a Josephson-like interference experiment.  At present, the best evidence of coherence is seen in the dramatic jump of tunneling current between the wells, which is analogous to superradiance \cite{tunnelingpeak}. 

These experiments, which use permanent excitons, can be seen as directly mapping to the case of optically generated excitons. In the standard bilayer experiment, there are nearly equal numbers of electrons and holes in each quantum well.  One can in principle tune the ratio of the number of electrons in one well relative to the number of electrons in the other while keeping the total number of electrons in the two wells constant. If the electrons were all in one well and the other well was empty, this would make the system look very much like a full valence band and an empty conduction band, where the two wells play the role of the two bands. Tunneling between the two wells then becomes analogous to optical transitions between the valence band and conduction band in a semiconductor, and enhanced tunneling analogous to superradiance. 

Having nearly all the electrons in one well would correspond very closely to the case of magnetoexcitons in Landau levels created by optical pumping, which can form a BEC state, studied in several works \cite{mosk}. In the actual experiments with bilayer excitons in magnetic field, when the ratio of electrons in the wells is tuned away from 50\% in each well, the dramatic Coulomb drag resonance effect disappears quickly \cite{kellogg}. The reason is not fully understood, but it may be that the repulsion of nearest neighbors at half filling plays a crucial role in forcing the carriers into correlated pairs. In this case of half-filled levels, the bilayer system is in many ways analogous to a high-density BCS state of optically generated excitons known as the excitonic insulator (EI), studied by Keldysh and others \cite{EI} as the BCS limit of the excitonic BEC-BCS crossover. 

\section{Conclusions}

Because of the effects of disorder even in the very pristine samples used in the optically-pumped coupled quantum well experiments, the most promising regime to look for coherent effects is most likely at high density, where disorder effects are screened out. In this regime the condensate fraction may be low, as in liquid helium, where the condensate fraction is 10\% or less. This may mean that the effects of condensation will be best observed in hydrodynamic effects in transport, rather than an optical spectroscopy signature such as a sharp peak in the luminescence spectrum. 

In the case of the experiments with coupled quantum wells with a permanent population of electrons in Landau levels in a magnetic field, evidence of condensation must come from transport experiments. Enhanced tunneling points to condensation, but it would be appealing to see a Josephson-like interference measurement, or possibly observation of quantized vortices. 

Despite the fact that the evidence presented for condensation in these systems is so far not conclusive, there are many nice properties of these systems which make them promising for condensation. Not least of these is that a moving condensate of indirect excitons in the plane would correspond to a superconductor. There is also much room for study of the basic physics of interacting dipoles in a two-dimensional system. 

{\bf Acknowledgements}. This work has been supported by the U.S. Department of Energy grant DE-FG02-99ER45780.

\end{document}